\documentclass{mn2e}
\usepackage{epsfig}
\begin{document}

\title[QSOs around NGC3079]{An anomalous concentration of QSOs around NGC3079}
\author[E.M. Burbidge ]
{E.M. Burbidge,$^{1}$
G. Burbidge$^{1}$, H.C. Arp$^{2}$ and W.M. Napier$^3$\\
$^{1}$ Center for Astrophysics and Space Sciences 0424 University of
California San Diego, CA  92093-0424, USA\\
$^{2}$ Max-Planck-Institut für Astrophysik 85741 Garching, Germany \\
$^{3}$ Centre for Astrobiology, Cardiff University, 2 North Road,
Cardiff CF10 3DY, Wales, UK }

\date{Last update 2005 September 03; in original form 2005 September 03}

\maketitle

\begin{abstract}

It is shown that there are at least 21 QSOs within 1$^\circ$ of the nearby active spiral
galaxy NGC3079. Many of them are bright (mag$<$18) so that the surface density of those
closer than 15$^\prime$ to the galaxy centre is close to 100 times the average in the
field. The probability that this is an accidental configuration is shown to be
$\la$10$^{-6}$. Discovery selection effects and microlensing fail by a large factor to
explain the phenomenon, suggesting that the QSOs lie in the same physical space as
NGC3079.  However, two of them make up the apparently lensed pair 0957+561A, B whose
lensing galaxy lies at $z$=0.355. This problem is discussed in the concluding section.
\end{abstract}


\section{Introduction}

NGC~3079 is a bright spiral galaxy (type Sc) which shows many signs
of activity in its central region. Several studies of it have been
made in optical, radio and X-ray energies (cf Filippenko and Sargent
1992; Pietsch et al 1998, Kondratko et al 2005) and it is generally
agreed that it is in the process of ejecting hot gas and plasma
clouds and streams which are emitting large fluxes of radio waves
and X-rays (cf Pietsch et al 1998). The most energetic known
wind-blown superbubble, with kinetic energy $\sim \! 2\times
10^{56}/n_e$~erg, emerges from its nucleus.

Three of us (EMB, GB, HCA), and others, have shown that there are
many other bright active galaxies similar to NGC~3079, around which
are found many QSOs with large redshifts, (for example Burbidge et
al 2003; Chu et al 1998; Arp et al 2001, 2002; Galianni et al 2005).
In those papers we concluded that the QSOs are very probably at the
same distances as the active galaxies, that they are physically
associated with them, and that they have been ejected from them.
Thus on this hypothesis the redshifts of these QSOs are of
non-cosmological origin.

In this paper we analyze the population of QSOs lying close to
NGC~3079, which is one of the nearest of the active galaxies. We
find that there is a highly significant excess of bright QSOs around
the galaxy, and that this excess cannot be explained by selection
effects or microlensing.

\section{Optical QSOs}

\begin{table*}
\begin{minipage}{126mm}
\caption{QSOs within 1$^\circ$ of NGC~3079. $F_x$ is in units of
$\rm 10^{-14}\,erg\,cm^{-2}\,s^{-1}$. That for 0957+561 refers to
both images. $d$ represents the angular distance from the nucleus of
the galaxy, $b$ the magnitude of the QSO.}
\begin{tabular}{lrrrcrc}
Object &  $\Delta\alpha\cos\delta$ & $\Delta\delta$ & $d$  & $b$ & $F_x$ & $z$   \\
       & $^\prime$ & $^\prime$ & $^\prime$  & &  &      \\ \hline
SBS0953+556         & -44.0 & -19.8 & 48.3  & 18.00 &     & 1.410 \\
4C55.17             & -36.7 & -17.9 & 40.8  & 17.40 &     & 0.900 \\
QO955+5623          & -28.2 &  28.4 & 40.0  & 18.31 &     & 0.066 \\
SBS0955+560         & -28.3 &  10.1 & 30.1  & 17.60 & 64  & 1.021 \\
RXJ10005+5536       & -12.2 &  -4.4 & 13.0  & 17.50 & 89  & 0.216 \\
1WGAJ1000.9+5541    &  -8.8 &  -0.2 & 8.8   & 19.10 & 19  & 1.037 \\
NGC3073UB1          &  -7.6 &   5.5 & 9.4   & 19.20 &     & 1.530 \\
ASV1                &  -7.0 &  12.3 & 14.1  & 17.28 &     & 0.072 \\
SBS0957+557         &  -6.8 & -12.2 & 14.0  & 17.60 & 20  & 2.102 \\
1WGAJ1000.3+554     &  -5.5 &   5.1 & 7.5   & 21.00 & 10  & 2.680 \\
0957+561A           &  -5.3 &  13.1 & 14.1  & 17.00 & 201 & 1.413 \\
0957+561B           &  -5.4 &  13.0 & 14.1  & 17.00 &     & 1.415 \\
ASV23               &  -4.9 &  10.9 & 12.0  &  --   &     & 0.900 \\
ASV24               &  -4.8 &  14.7 & 15.5  & 23.03 &     & 1.154 \\
ASV31               &  -4.4 &  14.6 & 15.3  & 21.14 &     & 0.352 \\
MARK132             &  -4.1 &  13.8 & 14.4  & 16.00 &     & 1.760 \\
QO958+5625          &  -2.2 &  29.7 & 29.8  & 20.08 &     & 3.216 \\
NGC3073UB4          &   1.1 &   2.1 & 2.3   & 18.10 &     & 1.154 \\
1WGAJ1002.7+5558    &   6.5 &  17.1 & 18.3  & 21.70 & 43  & 0.219 \\
1WGAJ1002.7+5541    &  10.0 &   0.7 & 10.0  & 18.50 & 15  & 0.673 \\
87GB100156.9+553816 &  24.0 & -17.0 & 32.0  &  --   & 65  & 0.431
\end{tabular}
\label{ta:QSO_list}
\end{minipage}
\end{table*}

From the compilation of QSOs with measured redshifts by Veron \&
Veron (2003), and  Guti\'{e}rrez (2005), we list in
Table~\ref{ta:QSO_list} twenty one QSOs which lie with $1^\circ$ of
the center of NGC~3079. Fig.~\ref{qad} shows that they are strongly
clustered in a highly asymmetrical way very near NGC~3079. In the
Veron catalogue references are given, showing how each QSO was first
identified.  It is clear from studying these papers and in some
cases from the names of the objects, that they have not been found
following a deliberate hunt for QSOs in that region.  For example,
one of them was found in the 4C radio survey, one is a Markarian
object with a very high redshift for that class, a number of them
were first identified as X-ray sources, later shown to be QSOs, two
were indeed found in looking for candidate QSOs near the adjacent
galaxy NGC 3073 (Mk 231), and of course the very close pair
0957+561A\&B was identified originally from a radio position and is
generally believed to be the prototype gravitational lens pair.

In the general field the best estimate of surface density of QSOs
suggests that there are no more than 0.3 per square degree for QSOs
brighter than 18.0 (Boyle et al 1991; Kilkenny et al 1997; Fig.~1 of
Myers et al 2005). But from Table~\ref{ta:QSO_list} we see that
there are seven QSOs with $m<18$ within 35$^\prime$ of the centre of
NGC~3079, six of them within 15$^\prime$ of the centre. Thus the
surface density of bright QSOs very close to NGC~3079 is close to
100 times the average for bright QSOs.

Similar high density clustering of QSOs had previously been found
about the Seyfert galaxy NGC 6212 and the QSO 3C345 (Arp 1997,
Burbidge 2003), though in that case the clustering tendency was seen
first, before it was realized that NGC 6212 is a very active
Seyfert.  NGC 6212 is further away than NGC~3079, and more QSOs are
present within $1^\circ$ of it.

According to Sandage and Tammann (1981) the distance of NGC~3079 is
$\sim$20.5 Mpc. If the QSOs are at the same distance as NGC~3079, 14
of them in Table~\ref{ta:QSO_list} lie within 100 kpc of the center
of the galaxy, eight of which are brighter than 18.5~mag. The most
distant QSO listed in the Table is still only about 300 kpc from the
center of the galaxy,

\section{X-ray emitting QSOs}
\begin{figure}
\begin{minipage}[b]{1.0\linewidth}
   \center{\includegraphics[angle=0,width=1.0\linewidth]{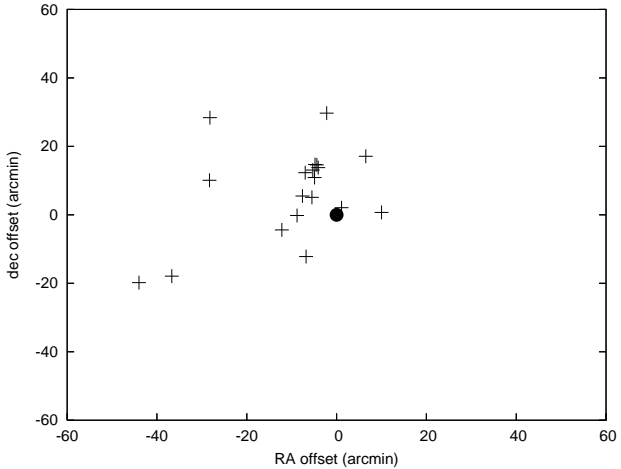}}
\end{minipage}
\caption{Distribution of QSOs within 1$^\circ$ of NGC~3079 (the
latter represented by the solid circle).} \label{qad}
\end{figure}

Of the 21 QSOs listed in Table~\ref{ta:QSO_list}, ten have been
identified as powerful X-ray sources (their fluxes are given in the
Table). The five most powerful x-ray emitters, with $F_x>50 \times
10^{-14}\rm erg\, cm^{-1}\, s^{-1}$, lie with about 36$^\prime$ of
NGC~3079 and are roughly aligned in the direction of the major axis
of the galaxy. They are shown in Fig.~\ref{qad}. The $\log N-\log S$
data from the RIXOS counts (Page et al 1996, their Fig.~3) reveal
that for $N =50 \times \rm 10^{-14} erg\, cm^{-1}\, s^{-1}$, the
surface density in the field is about $\rm 0.1\,deg^{-2}$. However,
in the area of about $\rm 1.1\,deg^2$ we find four sources (or five
if we count 0957+561 A\&B as two) where we should only see about
0.12 sources.

There are 7 X-ray emitting QSOs lying within 15$^\prime$ of
NGC~3079, with $F_x \ga 10 \times \rm 10^{-14} erg\, cm^{-1}\,
s^{-1}$ corresponding to a surface density of about 30 per square
degree. From the results of Hasinger et al (1998), the density of
such sources in the field is about $\rm 3.2\,deg^{-2}$ (their
Fig~4b), and once again we find a density about ten times that
expected. Thus we find the same effect for the X-ray QSOs as we do
for the optically bright ones, although as the X-ray QSOs tend also
to be bright in this sample, the datasets are not independent.

\section{Statistical significance of the clustering}

\begin{figure}
\begin{minipage}[b]{1.0\linewidth}
   \center{\includegraphics[angle=0,width=1.0\linewidth]{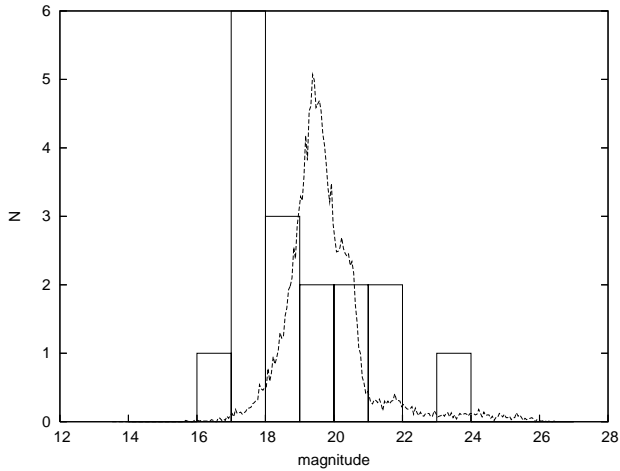}}
\end{minipage}
\caption{Magnitude distribution of QSOs within 1$^\circ$ of
NGC~3079, compared with that of QSOs in the Sloan survey
(represented by the dotted line, normalised).} \label{magdis}
\end{figure}

\begin{figure}
\begin{minipage}[b]{1.0\linewidth}
   \center{\includegraphics[angle=0,width=1.0\linewidth]{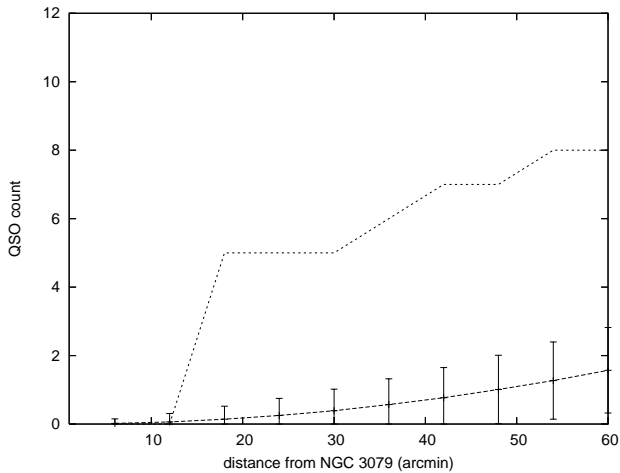}}
\end{minipage}
\caption{Upper curve: cumulative distribution of QSOs $b<$18.0
around NGC~3079. Lower curve: expected background count assuming
100\% completeness of discovery. Error bars are 1$\sigma$ limits
based on Poissonian estimates.} \label{xs18}
\end{figure}

\begin{figure}
\begin{minipage}[b]{1.0\linewidth}
   \center{\includegraphics[angle=0,width=1.0\linewidth]{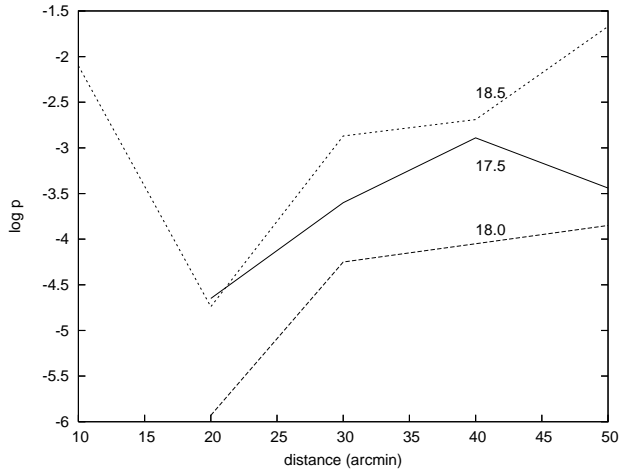}}
\end{minipage}
\caption{Chance probabilities of finding QSOs brighter than $b_t$ as
a function of angular distance from NGC~3079. These are based on an
assumed 100\% discovery of QSOs and so are a severe upper limit. The
curves are labelled with the threshold magnitudes.} \label{mp}
\end{figure}

\begin{table}
\caption{Cumulative number of QSOs with magnitudes $m\la 18.5$~mag within angular
distance $\theta^\prime$ of NGC~3079, compared with expectations from background count
$n_b$, counting errors $\surd n_b$.}
\begin{tabular}{rcrr}
$\theta$ &   $n(\la\theta)$  &  $n_b$  &  $\surd n_b$  \\ \hline
 6       &   1              &   0.06  &   0.24        \\
12       &   2              &   0.23  &   0.48        \\
18       &   7              &   0.51  &   0.71        \\
24       &   7              &   0.90  &   0.95        \\
30       &   7              &   1.41  &   1.19        \\
36       &   8              &   2.04  &   1.43        \\
42       &  10              &   2.77  &   1.66        \\
48       &  10              &   3.62  &   1.90        \\
54       &  11              &   4.58  &   2.14        \\
60       &  11              &   5.65  &   2.38        \\
\end{tabular}
\label{ta:ang_dist}
\end{table}

In Fig.~\ref{magdis} the magnitude distribution of the QSOs around
NGC~3079 is compared with that from the Sloan survey. It is clear
that they are systematically brighter than QSOs in the general
field. A $t$-test on the two samples reveals that the difference
remains highly significant ($>99$\% confidence level) when the
magnitude distributions are cut off at $b$=20, and significant
(93\%) at $b$=19.

As we have seen most of the QSOs around NGC~3079 were discovered as
part of radio or x-ray surveys rather than due to selective scrutiny
of the neighbourhood of the galaxy. However, even if the latter were
the case, this would not account for the excess since the number
density of bright background QSOs is too low. This can be seen from
Table~\ref{ta:ang_dist} and also Fig.~\ref{xs18}.

Poisson statistics can be used to put formal probabilities on the
concentration of bright QSOs around NGC~3079, as a function of
threshold magnitude $b_t$ and angular distance $\theta$ from the
nucleus of the galaxy. The results are shown in Fig.~\ref{mp}, and
confirm that the concentration is very unlikely to be due to chance.
In particular the presence of four QSOs brighter than magnitude 18
within 20$^\prime$ of the nucleus has a chance probability of about
one in a million. The calculation is made on the assumption that the
list of bright QSOs in Table~\ref{ta:QSO_list} is 100\% complete,
which is probably not the case, and so the probabilities represent
upper limits. Of course a posteriori statistics will readily yield
low probabilities for coincidences of no true physical significance;
however the present probabilities should probably not be put in this
category since there are several pre-existing claims in the
literature for QSO excesses around many bright, nearby, active
spirals (loc. cit.); thus a QSO excess is reasonably seen as an a
priori hypothesis to be tested. A completely rigorous test for a
physical association would require a full sampling of such galaxies
down to a prescribed magnitude limit and range of morphologies.

An alternative hypothesis to be considered is that NGC 3079 lies
fortuitously close to an unexplained concentration of bright QSOs
but otherwise has no connection with it. However the galaxy is
amongst the brightest 150 spirals in the sky, only a few percent of
which are active and therefore candidates for testing the hypothesis
of QSO/active spiral correlation. If the mean separation between
such galaxies is say $45^\circ$, chance proximity of one to the QSO
concentration has probability $(0.25/45)^2\sim 3\times 10^{-5}$.

\section{Microlensing}

A background QSO may be microlensed if it lies along the line of
sight of foreground sources in the halo of NGC~3079 to within a few
Einstein radii. The Einstein ring has angular radius, in arcseconds,
\begin{equation}
\theta_E = 3\sqrt{m/D}
\end{equation}
where $m$ is the mass of the lens in solar masses and $D$ is its
distance in Gpc (Narayan \& Bartelmann 1996). If there are $n$ such
lenses in the halo of NGC~3079, whose distance $D\sim$0.02~Gpc, then
the total lensed area is, in $\mu$arcsec,
\begin{equation}
A=n\pi \theta_E^2 = 9\pi M/D = 67 M^{1/2}
\end{equation}
from eqn (1), where $M=nm$ is the total mass of lensing objects in
the halo. Thus the proportion of QSOs lying within an Einstein ring
is independent of the adopted lensing parameters (Narayan \&
Bartelmann 1996).

For lensing masses totalling $M=10^{12}$~M$_\odot$, the total lensed
area is equivalent to a single disc of Einstein radius $\Theta
\sim$21~arcsec and area $\sim 10^{-4}$ square degrees. For
comparison, there are about 20 QSOs per square degree down to
magnitude 20 and 100 down to magnitude 22 (Myers et al. 2005). Thus
with these figures there is an expectation that $\sim$0.002
background QSOs with $b<$20 lie within Einstein rings in the halo,
and $\sim$0.01 QSOs with $b<22$.

A microlensed QSO an angular distance $\theta = \theta_E$ from a
lens is brightened by $\delta b\sim0.2$~mag, one within
0.1$\theta_E$ by up to two magnitudes. Assume that the 12 QSOs
around NGC~3079 with $b<$18.5 are for the most part microlensed,
being enhanced above background number density by a factor $\sim$10,
say. The number density of QSOs to $b=20$ is about ten times that
down to $b=18.5$, but since there are only 0.002 QSOs with $b<$20
available for magnification within the Einstein rings, these QSOs
are about four powers of ten short of supplying the excess. For
background QSOs with $b<22$ the $\delta b$ required is now 3.5 or
more, requiring $\theta \la 0.1 \theta_E$, and the area available to
supply this degree of lensing is $\sim\! 10^{-6}$~sq~deg. With
background number density $\sim$100 per square degree, QSOs with
$b<22$ are five powers of ten short of yielding a factor 10 above
background of the bright QSOs around the galaxy. Thus as one
attempts to boost progressively fainter background QSOs,
microlensing increasingly fails to account for the observed excess
around NGC~3079 (the argument is conservative since it neglects the
lowering of density of background QSOs caused by the magnification).
This deficiency arises from the increasing flatness of the slope of
the QSO number-magnitude distribution with increasing $b$; an
increase in the QSO count is expected behind lenses when the slope
of the QSO number-magnitude count $>$0.4, a deficit otherwise, with
the break-even point occurring around magnitude 19.1 -- 19.6 (Myers
et al. 2005). The failure of lensing to account for the excess of
QSOs around galaxies generally on sub-Mpc scales had previously been
pointed out by Arp (1990).

The galaxy between the archetypal lensed image 0957+561A\&B lies
near the centre of a cluster of redshift $z = 0.355$, mass $\sim
3.5\times 10^{14}$~M$_\odot$ and angular radius $\sim$2~arcmin
(Angonin-Willaime et al. 1994, Fischer et al. 1997). Remarkably,
four other QSOs lie remarkably close to 0957+561: the angular
distances of (ASV23, ASV24, ASV31, MARK132) from it are respectively
(2.36, 1.35, 0.69, 0.74) arcmin, which also places them, angularly,
within or close to this cluster of galaxies. The QSOs have
magnitudes \linebreak (---, 23.03, 21.14, 16.00) respectively. The
probability that they lie within one or two arc minutes of 0957+561
and each other by chance is $\sim$10$^{-5}$.

In this case the optical depth of the cluster to microlensing is of
order unity, but with an area $\sim\! 3\times 10^{-3}$~sq~deg there
is an expectation of only $\sim$0.3 QSOs with $b<$22 lying behind
it, and arguments similar to those presented above reveal that
microlensing within the $z = 0.355$ galaxy cluster is also incapable
of producing this degree of concentration of bright QSOs. For
multiple imaging, typical angular separations are $2\theta_E$. Apart
from 0957+561, there is no sign of multiple imaging or arclets in
the sample. Thus, within 14 arc minutes of NGC~3079, we find what
seems to be a second `anomaly': a group of bright QSOs whose close
proximity to each other is not readily explicable in terms of
conventional lensing.

If this group of QSOs, including 0957+561, is unconnected with
NGC~3079, then the chance probabilities computed in the previous
Section are affected. However only two of them are brighter than
magnitude 18.5, and the clustering of the remaining galaxies with
NGC~3079 has still $\la 10^{-4}$ probability of being due to chance.

\section{Discussion}

A straightforward analysis of the numbers of optical and X-ray QSOs
around the highly active galaxy NGC~3079 reveals that there is a
high density of these objects very near to the galaxy. For the most
luminous ones, which in all surveys of QSOs are very rare, the
density of QSOs near to the galaxy is about two orders of magnitude
higher than the density in the field. We have shown that this excess
cannot be accounted for by selective searches for QSOs around the
galaxy, and that the probability of accidental concentration is $\la
10^{-6}$, or $\la 10^{-4}$ if we -- somewhat arbitrarily -- exclude
the little group close to 0957+561. The close mutual proximity of
this little group itself seems to be anomalous. The active nature of
the galaxy corresponds to that around which other QSO concentrations
have been claimed.

If not due to an extraordinary chance, the concentration of bright
QSOs around NGC~3079 must arise either because they are physically
associated with it, or because they are background objects whose
brightness is amplified by gravitational lensing or microlensing by
condensed objects in the halo of the galaxy. The former hypothesis
is discussed by Arp et al (2005), who describe evidence from the
radio and x-ray studies that some of the QSOs and the nucleus of
NGC~3079 may be physically associated. Here we have confined
ourselves to exploring the lensing hypothesis. We have found that,
at least with conventional scenarios, lensing fails quantitatively
to account for the phenomenon. A similar problem has been met in
numerous other studies of QSO concentration around galaxies (e.g.
Kaiser 1992, Rodrigues-Williams \& Hogan 1994). Myers et al (2005),
in a comprehensive study of statistical lensing involving 2df QSOs,
find correlations and anti-correlations consistent with the
predictions of statistical lensing, except that on 100 kpc scales --
those considered here -- the results require more mass present than
is predicted by current $\rm \Lambda CDM$ models.

This might argue for the QSO ejection hypothesis by default. However
there is a problem here too: one of the QSOs is the famous pair
0957+561 A\&B, considered by many to be the prototype
gravitationally lensed pair. In this case, the lensing galaxy has
been identified, and a good model for the system has been made. If
the QSO is indeed gravitationally lensed, it would presumably be a
background object. As far as we are aware only Weymann (1995) has
ever pointed out before that 0957+561A,B lies so close to NGC~3079.
However the concentration of QSOs very close to this lensed pair is
also anomalously high. It is therefore conceivable that these
excesses, occurring around both NGC~3079 and the cluster of galaxies
aligned with 0957+561, are manifestations of the same phenomenon.

\end{document}